Commentary on "Is coding a relevant metaphor for the brain?" (Brette 2019). To appear in *Behavioral and Brain Sciences*.

# Codes, communication and cognition

Stevan Harnad
Department of Psychology
Université du Québec à Montréal
and
Department of Electronics and Computer Science
University of Southamton
harnad@soton.ac.uk

**Abstract:** Brette criticizes the notion of neural coding because it seems to entail that neural signals need to be "decoded" by or for some receiver in the head. If that were so, then neural coding would indeed be homuncular (Brette calls it "dualistic"), requiring an entity to decipher the code. But I think Brette's plea to think instead in terms of complex, interactive causal throughput is preaching to the converted. Turing (not Shannon) has already shown the way. In any case, the metaphor of neural coding has little to do with the symbol grounding problem.

Both Shannon's (1948) *information* and Turing's (1936) *computation* are important in cognitive science. Shannon is concerned with the faithfulness of signal transmission in communication and Turing is concerned with what algorithms can do. Cognitive science is concerned with what organisms (hence their brains) can *do*, and *how*.

Cells (including neurons) transmit signals. This is already true in plants (Baluska & Mancuso 2009) and of course also in machines. And organisms certainly do things. Which of the things organisms do are "cognitive" and which are "vegetative" is mostly just a definitional matter, but it is probably overstretching the notion to say that paramecia or hearts are "cognizing." The examples are nevertheless instructive for cognitive science, because paramecia, hearts and organisms with brains are all systems that can *do* things. So are computers and robots, for that matter. Hence finding a causal explanation of how one of them does what it does may provide useful lessons for explaining the others.

Let's start with the heart, an example used by Brette. What does the heart do? It pumps blood. No metaphors. The heart literally pumps blood, and cardiac science has successfully reverse-engineered the heart (to a close approximation). We know how the heart does it -- and part of the proof that we know how is that we can apply and test our hypotheses about how the heart pumps blood by building a synthetic model of a heart, plugging it into the heart's inputs and outputs, and testing whether it can pump blood. If it can, the artificial heart passes the "Turing Test" for cardiac function.

So what does the (human) brain (and body) pump? Human behavior. Or, rather, human behavioral *capacity*. What people *can do*. Let's forget about what portion of that capacity counts as cognitive and what proportion is just vegetative (like cardiac function): It all consists of the capacity of a (living) system to do certain things. Now the challenge is to explain how.

Turing (1950) provided the ground rules: You have an explanation if you can design a system that can do everything a human being can do, indistinguishably -- to a human -- from a human. If your interest is just in "cognitive" capacities, then just generate those, ignoring the vegetative capacities (or at least those that are not essential for generating the cognitive capacities). Cognition, like Justice Potter Stewart's pornography, may be hard to define, but we know it when we see it. And the capacity to interact with the dynamic world of objects and events and their properties (including words describing those objects, events and properties) indistinguishably from the way humans do, is surely cognitive, if anything is.

There is one more thing: Humans don't just do: They also feel. It *feels like something,* to a human, to be seeing and doing what humans can see and do. But the capacity to feel eludes Turing's program for cognitive science. It's something our brains pump invisibly. Turing (1950) accordingly brackets it. But it keeps making disruptive peekaboo appearances in our attempts to reverse-engineer cognition, as we shall see.

One of the main hypotheses about how the brain pumps cognitive capacity is via computation, Turing computation. Computation is the manipulation of "symbols" (arbitrary formal objects) on the basis of rules operating only on the symbols' shapes ("syntax"), not their meanings ("semantics"), in order to generate certain symbolic outputs from certain symbolic inputs. That's what algorithms do. (An intuitive example is the rule we all learned in school for extracting the roots of quadratic questions: "minus b plus or minus the square root of…".)

Algorithms are like recipes: apply them to the symbolic ingredients and you can explain how to bake a symbolic cake. Computation is very powerful; just about everything in the universe can be encoded symbolically and explained computationally, including cardiac function. The right algorithm can pump symbolic blood. And you can show that the algorithm really works by applying it to build a synthetic heart that really passes the cardiac Turing Test and pumps blood. But to do that, you have to "interpret" the symbolic code and implement it in material form, just as a formal recipe for a cake needs to be implemented in material form, using the real ingredients referred to by the symbols, in order to generate a real cake.

So, despite its enormous power, computation cannot be all there is to cognition. Searle (1980) showed, famously (in this journal), that a computer is not cognizing even if it can pass the Turing Test (TT) because Searle too could pass the Chinese TT by executing the symbolic code without understanding a word of Chinese. Why can't he understand? Because there is no connection between the symbols in the code and the objects in the world that they are interpretable as being about. Interpretable by whom? The user or the executor of the code. But the meaning itself is not in the code.

That is the symbol grounding problem (Harnad 2006). Simple solution: The TT must not be merely symbolic (verbal). It must test not only what the candidate can say, but also all the other things a human cognizer can do in interacting with the objects in the world that the verbal TT is merely chatting about. The candidate has to be a robot. And a Turing robot is not just a computer, manipulating formal symbols; it is a dynamical system, able to interact with the objects in the world. Its symbols are grounded in its capacity to identify and interact with their referents indistinguishably from the way we do.

Now to neural "codes":  Brette is right that it would be homuncular (although he calls it "dualistic") to think of input to sensory receptors, activity along sensory pathways to sensory and sensorimotor regions in the brain -- and then onward to motor regions and pathways to motor effectors – as encoded signals being transmitted in order to be decoded by a receiver, as in telegraphic communication of morse code from a sender to a receiver. There is no homunculus on the receiving end. It's all just a dynamic causal process constituting the organism's capacity to do what it can do, some of it output in response to immediate sensory input, some of it generated by endogenous processes.

But it is harmless to call the neural activity along sensory input pathways a "neural code." Shannon's communication theory is about the end-to-end fidelity of signal transmission (of analog or digital signals); it is not about cryptography, let alone about the interpretation of computational algorithms or of natural language. To show that there is a substantive issue involved here, Brette would have to show that there is a nontrivial chunk of performance capacity (even  the detection of interaural time difference: ITDs) that cannot be explained causally if we insist on calling the activity occurring along the sensory input pathways a "neural code." (Brette's preferred notion of "neural representations," by the way, sounds just as homuncular to me as the idea of neural codes: "representation of what, to whom?" Ditto for "internal model.")

Let me close with Brette's fleeting mention of "percepts." This is an instance of the "peekaboo" influence of homuncular thinking. Psychophysics, too, can only study what the organism does (input/output), not whether or how it feels like something to do it. Sensorimotor activity is only perceptual if it is felt. I don't doubt that it feels like something to detect an IDT, just as it feels like something to understand Chinese. But although symbol-grounding and Turing-testing may be "easy" (in principle, if not in practice), explaining how and why organisms *feel* rather than just *do* is and remains notoriously hard.

**References**

Baluška, F., & Mancuso, S. (2009). Plant neurobiology: from sensory biology, via plant communication, to social plant behavior. *Cognitive Processing*, *10*(1), 3-7.
Brette R. (2019) Is coding a relevant metaphor for the brain? *Behavioral and Brain Sciences*. February 2019 1–44.
Harnad, S. (2006). The symbol grounding problem. *Encyclopedia of Cognitive Science. Wiley*

4Harnad, S. (2009[) The Annotation Game: On Turing (1950) on Computing, Machinery, and Intelligence](). In: Epstein, R., Roberts, G., & Beber, G. (Eds.). *Parsing the Turing test*. Springer Netherlands.

Shannon CE (1948) A mathematical theory of communication. *Bell system technical journal*, *27*(3): 379-423.

Turing, A. M. (1936). On computable numbers, with an application to the Entscheidungsproblem. *Proceedings of the London mathematical society*, *2*(1), 230-265.

Turing, A. M. (1950) Computing Machinery and Intelligence. *Mind 49:* 433-460.

Harnad, S. (2009) The Annotation Game: On Turing (1950) on Computing, Machinery, and Intelligence. In: Epstein, R., Roberts, G., & Beber, G. (Eds.). *Parsing the Turing test*. Springer Netherlands.

Shannon CE (1948) A mathematical theory of communication. *Bell system technical journal*, *27*(3): 379-423.

Turing, A. M. (1936). On computable numbers, with an application to the Entscheidungsproblem. *Proceedings of the London mathematical society*, *2*(1), 230-265.

Turing , A. M. (1950) Computing Machinery and Intelligence. *Mind 49:* 433-460.